# Training of mixed-signal optical convolutional neural network with reduced quantization level

Joseph Ulseth, Zheyuan Zhu, Guifang Li, and Shuo Pang

CREOL, The College of Optics and Photonics, University of Central Florida, 4304 Scorpius St, Orlando, FL, 32816

## Abstract

Mixed-signal artificial neural networks (ANNs) that employ analog matrix-multiplication accelerators can achieve higher speed and improved power efficiency. Though analog computing is known to be susceptible to noise and device imperfections, various analog computing paradigms have been considered as promising solutions to address the growing computing demand in machine learning applications, thanks to the robustness of ANNs. This robustness has been explored in low-precision, fixed-point ANN models, which have proven successful on compressing ANN model size on digital computers. However, these promising results and network training algorithms cannot be easily migrated to analog accelerators. The reason is that digital computers typically carry intermediate results with higher bit width, though the inputs and weights of each ANN layers are of low bit width; while the analog intermediate results have low precision, analogous to digital signals with a reduced quantization level. Here we report a training method for mixed-signal ANN with two types of errors in its analog signals, random noise, and deterministic errors (distortions). The results showed that mixed-signal ANNs trained with our proposed method can achieve an equivalent classification accuracy with noise level up to 50% of the ideal quantization step size. We have demonstrated this training method on a mixed-signal optical convolutional neural network based on diffractive optics.

## 1. Introduction

Artificial neural networks (ANN) are growing larger and deeper [1–3] to tackle tasks of increasing complexity [4–6]. To accommodate the computation demand in future neural network structures, specialized computing hardware and data formats have been engineered. Various low-precision or even binary neural networks (BNNs) accompanied by specifically designed training algorithms [7–10] have proven successful in accelerating the inference and reducing the memory footprint [11,12] by using low-bit width, fixed-point data format for the weights and inputs. When designing and deploying these networks on digital computers [10], intermediate results (e.g. activations) often need to be cached in a higher precision format than the weights and inputs to achieve the expected accuracy.

Recently, due to the advantages in speed and power efficiency [13], analog computing paradigms have been considered as solutions to the growing demand in neural network computing, with implementations in both electronics [14,15] and photonics [16,17]. However, analog computing is susceptible to ambient noise and device imperfections [18]. Ex-situ training has been deployed on a simulated analog unit using fixed-point data format [19], analogous to the low-bit width neural network using a digital computer. Yet a model trained by such method is likely to have an inferior inference performance [20], as analog intermediate results cannot match the full precision in a digital computer. To overcome this performance degradation, fine-tuning of the analog parameters on each computation node [15,21] can be performed, though requiring an exhaustive effort in training. There has not been an efficient training method that is robust to the errors on analog ANNs.

In this work, we incorporate two types of common analog computation errors – random noise and deterministic errors – into the training process, extending low-precision neural networks training to mixed-signal or analog computing platforms. The network trained with our method is robust against analog signal noise level as high as 50% of the quantization step, indicating that mixed-signal neural network can operate at a reduced quantization level. We have demonstrated a trained model on a programmable optical convolutional neural network.

## 2. Theory

*2.1 Low-precision training*

Low-precision neural networks perform matrix multiplications or convolutions between fixed-point inputs and weights, which are typically the required data format for many digital tensor processing units [22]. Fig. 1(a) illustrates the computation scheme of a low-precision neural network layer. A fixed-point processor computes the activations $h^{(k)}$ from the quantized inputs $a^{(k)}$ and weights $W^{(k)}$

$$h^{(k)} = a^{(k-1)} \cdot W^{(k)}, \tag{1}$$

where $\cdot$ denotes the matrix multiplication or convolution operation. The activations $h^{(k)}$ require higher bit width than the inputs and weights due to the associated accumulation process [10]. A nonlinear function $g(\cdot)$ is then applied on the activations, along with a quantization operation, to match the input data format of the next layer,

$$a^{(k)} = quantize(g(h^{(k)})). \tag{2}$$

Here $g(\cdot)$ can be a common neural network operation such as batch normalization, down-sampling, ReLU etc., or a combination of multiple operations. The quantization operation is defined as,

$$quantize(x) = \begin{cases} \left[\frac{x}{2^{L-m-1}}\right] 2^{L-m-1}, |x| < (2^m - 2^{m+1-L}) \\ sign(x)(2^m - 2^{m+1-L}), |x| \geq (2^m - 2^{m+1-L}) \end{cases}, \tag{3}$$

where $L$ is the total bit width, $m$ is the number of integer bits, $[\cdot]$ denotes the rounding to the nearest integer. Several nonlinearities, such as clipping and scaling functions, have been purposefully designed for easier integration with the quantization operation [23,24].

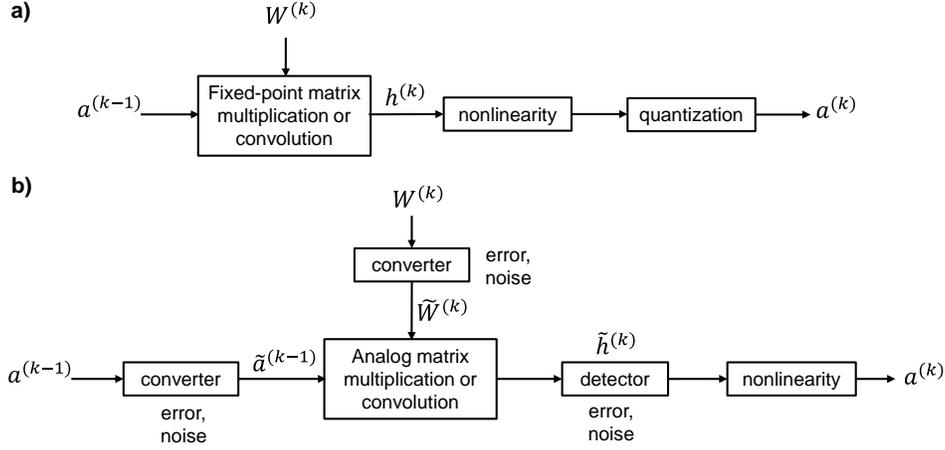

Figure 1: Computation scheme of (a) digital fixed-point neural network layer (b) a neural network layer with an analog acceleration unit. Variables marked by tildes are analog signals.

*2.2 Mixed-signal ANN layer with an analog acceleration unit*

A growing number of neural network architectures have replaced the traditional digital fixed-point matrix multiplication or convolution calculations with the analog counterparts for speed and power efficiency [15–17]. Fig. 1 (b) illustrates the computation scheme of a mixed-signal neural network layer with an analog acceleration unit. A set of digital inputs $a^{(k-1)}$ and weights $W^{(k)}$ are sent to their corresponding digital-to-analog convertors (DAC), generating the inputs $\tilde{a}^{(k-1)}$ and $\widetilde{W}^{(k)}$, respectively, for analog accelerators. The activations $\tilde{a}^{(k-1)} \cdot \widetilde{W}^{(k)}$ from the analog acceleration unit are collected by a detector, producing signals $\tilde{h}^{(k)}$. The detected signals are then sent to a digital processing unit, which maps the activations to the inputs of the next layer via a nonlinear function, $a^{(k)} = g(\tilde{h}^{(k)})$, that can include similar operations as those in digital low-precision neural networks.

Computational errors in an analog acceleration unit include random noises or deterministic errors. These two types of errors can apply to any analog signals $\tilde{a}^{(k-1)}$, $\widetilde{W}^{(k)}$, and $\tilde{h}^{(k)}$. From the statistics perspective, random noises introduce a variance to the signal, and the deterministic errors introduce a bias to the signal.

Deterministic errors typically originate from the nonlinear response $f(\cdot)$ of the modulators, DACs, or the detectors. The output of $f(\cdot)$ can be either continuous or discrete. For a continuous output, the signal $\tilde{x}$ distorted from the ideal signal $x$ is given by

$$\tilde{x} = f(x). \tag{4}$$

Examples of continuous deterministic error include gamma curves of the detector, or the sinusoidal relation between the intensity and phase in an interferometry-based intensity modulator [25]. The discrete deterministic error maps tensor $x$ to a set of values determined by hardware specifications

$$\tilde{x} = f(x) = \underset{x' \in \mathbf{X}'}{\mathrm{argmin}} \, |x' - x|_1. \tag{5}$$

where $\mathbf{X}'$ is the set of discrete values. Examples of discrete deterministic error include DACs that can only generate discrete voltage levels, or analog-to-digital convertors (ADCs) that digitize an analog signal with a fixed number of levels. The quantization error in digital low-precision or binary networks can be considered as special cases of Eq. (5), in which the set $\mathbf{X}' = \{\pm i \cdot 2^{m+1-L}; i = 0, 1, \ldots, 2^{L-1} - 1\}$ for fixed-point quantization with $L$ bits.

Noise is modeled as a random variable added to an ideal signal $x$. The signal corrupted by noise $\tilde{x}$ can be expressed by

$$\tilde{x} = x + \varepsilon, \tag{6}$$

where $\varepsilon$ is assumed to follow an unbiased distribution. If the random noise in an experimental platform introduces a bias to the signal, this bias can be merged into the fixed-pattern distortion. Notice that random noise and deterministic errors can be combined to model the errors associated with any analog signals in a practical analog acceleration unit.

## 2.3 Training of neural networks with analog computation units

Our proposed training method for mixed-signal ANN considers the two types of errors described above in both the forward pass and the gradient backpropagation during the training process. The gradient flow through Eq. (6) can be computed from the noisy instance of the tensor used in the forward inference [26]

$$\frac{\partial l}{\partial x} = \frac{\partial l}{\partial \tilde{x}}. \tag{7}$$

The gradient flow through the deterministic error process $\tilde{x} = f(x)$ (Eq. (4)) involves the derivate of $f(\cdot)$

$$\frac{\partial l}{\partial x} = \frac{\partial l}{\partial \tilde{x}} f'(x). \tag{8}$$

The derivate of a continuous nonlinear response $f(\cdot)$ is readily available. In the case that the output of $f(\cdot)$ is discrete, the gradient is 0 almost everywhere since $f(\cdot)$ is piecewise constant. To preserve the gradient flow, we use a gradient clipping method similar to that in BNN [8]

$$\frac{\partial l}{\partial x} = \frac{\partial l}{\partial \tilde{x}} \mathbf{1}_{\theta_1 < x < \theta_2}, \tag{9}$$

where $\partial l / \partial \tilde{x}$ is the gradient with respect to the distorted tensor $\tilde{x}$; $\mathbf{1}_{\theta_1 < x < \theta_2}$ denotes a binary tensor with the same shape as the ideal tensor $x$, with value 1 for elements of $x$ within the range $(\theta_1, \theta_2)$, and 0 for elements of $x$ outside the range $(\theta_1, \theta_2)$. $\theta_1$ and $\theta_2$ are typically chosen as the output range of $f(\cdot)$. For the special case of binarization, $\theta_1$ and $\theta_2$ are -1 and 1 (or 0 and 1 if $\mathbf{X}' = \{0,1\}$ in Eq. **Error! Reference source not found.**), respectively.

# 3. Mixed-signal convolution neural network simulation

## 3.1 Network structure

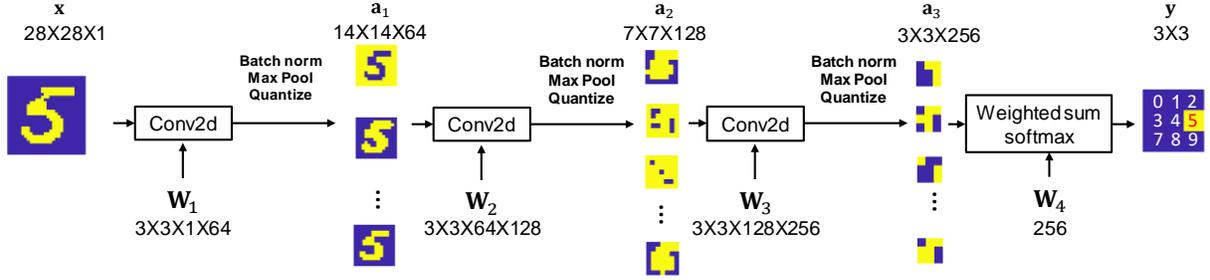

Figure 2: Structure of the convolutional neural network in the simulation.

We have constructed a mixed-signal, low-precision convolutional neural network, termed MCNN, that classifies the input digit, as shown in Fig. 2, with binary inputs and kernels in all layers. The MCNN consists only of convolutional layers to facilitate its deployment on a mixed-signal diffractive-optics-based system later. The input digit from the MNIST dataset (28X28) is gradually down-sampled to a 3×3 image representing the probability of each digit given the input image. The first layer convolves the 28×28 input image with 64 3×3 kernels, and outputs a 64-channel activation tensor with a size of 28×28×64. The 64 channels are then individually batch-normalized and max-pooled with 2×2 down-sampling to a 14X14X64 tensor as the input of layer 2. Layer 2 and 3 consist of the same convolution and post-processing operations, except that the number of kernels used are 128 and 256, respectively. The input of layer 4 is a 3×3×256 tensor down sampled from layer 3 activations by extracting the 2nd, 5th and 7th element along the horizontal and vertical spatial dimensions. Layer 4 performs a weighted sum of all 256 channels, applies softmax activation function, and outputs a final 3X3 image. Because only 9 possible labels can be produced from this MCNN, we excluded the digit '6'.

The computation errors that we consider in this MCNN simulation are the discrete deterministic errors on the inputs and weights, as well as random noise on the detector. In layer $k$, the analog inputs $\tilde{a}^{(k)}$ and weights $\widetilde{W}^{(k)}$ produced by the convertors from the ideal, digital values $a^{(k)}$ and $W^{(k)}$ are

$$\tilde{a}^{(k)} = f_a\big(a^{(k)}\big) = \operatorname*{argmin}_{a' \in \mathbf{A'}} |a' - a^{(k)}|_1,$$
$$\widetilde{W}^{(k)} = f_W\big(W^{(k)}\big) = \operatorname*{argmin}_{W' \in \mathbf{W'}} |W' - W^{(k)}|_1, \tag{10}$$

respectively. Here $\mathbf{A'}$ and $\mathbf{W'}$ are the sets of discrete input and weight tensors. For an input with M×N pixels and $Q$ input channels, the activations in a CNN convolution with 3X3 kernels and $C$ output channels are computed by an analog accelerator as

$$\tilde{h}_{i,j}^{(k,q,c)} = \sum_{m=i-1}^{i+1} \sum_{n=j-1}^{j+1} \tilde{a}_{m,n}^{(k-1,q)} \widetilde{W}_{m-i+2,n-j+2}^{(k,q,c)} + \varepsilon \tag{11}$$

where $i, j$ denote the index of the convolutional result; $m, n$ are the index of the input image; $q$ is the index of input channels, and $c$ is the index of output channels; $\varepsilon$ denotes the random additive noise, which is modeled by an unbiased Gaussian distribution $\varepsilon \sim \mathcal{N}(0, \sigma^2)$. Here we assume that the inputs are zero-padded. For simplicity, we omit the pixel and channel indexes in the tensor when they are not ambiguous. The activations $\tilde{h}^{(k,q,c)}$ then undergoes digital post-processing, which consists of batch normalization and 2X2 max pooling,

$$a^{(k)} = g(\tilde{h}^{(k,q,c)}) = MaxPool\left(BatchNorm\left(S(\tilde{h}^{(k,q,c)})\right)\right). \tag{12}$$

Here $S$ represents summation over all input channels $q$ of $\tilde{h}^{(k,q,c)}$ for each applied kernel $c$

$$\tilde{H}^{(k,c)} = S(\tilde{h}^{(k)}) = \sum_{q=1}^{Q_k} \tilde{h}^{(k,q,c)}, \tag{13}$$

and $BatchNorm(\cdot)$ is performed channel-wise on $\tilde{H}^{(k,c)}$ as

$$BatchNorm(\tilde{H}^{(k,c)}) = \gamma^{(k,c)} \left(\frac{\tilde{H}^{(k,c)} - \mu^{(k,c)}}{\sigma^{(k,c)}}\right), \tag{14}$$

where $\mu^{(k,c)}$, $\sigma^{(k,c)}$ are the mean and standard deviation among all pixels in the aggregate intermediate $\tilde{H}^{(k,c)}$ respectively, in channel $c$, and $\gamma^{(k,c)}$ is a trainable parameter. After $BatchNorm$, the features are binarized to obtain the input of the next layer.

### 3.2 Training of MCNN

The random noises and deterministic errors are both quantified by the root-mean-square-error (RMSE), which indicates the average deviation per dimension of the tensor

$$RMSE = \sqrt{\frac{|\tilde{x} - x|_2^2}{\dim(x)}}. \tag{15}$$

Here $\tilde{x}$ is the tensor $x$ corrupted by errors; $\dim(x)$ is the total number of elements in tensor $x$; $|\cdot|_2$ denotes the L2-norm. If $\tilde{x}$ is corrupted by unbiased Gaussian noise $\mathcal{N}(0,\sigma^2)$, the RMSE reduces to the standard deviation, $\sigma$.

The MCNN was trained considering binary inputs $\mathbf{A}' = \{0,1\}$, the kernel sets $\mathbf{W}'$ that can be displayed in the experiment, and a noise level $\sigma$=0.5 of the parameter $\varepsilon$. These parameters were selected to match the experimental MCNN setup. As a comparison, we also trained a reference MCNN model of the same structure, but without the random noise term $\varepsilon$ as in Eq. (11). The training of the reference model was similar to the BNN training [8], except that the binarization on the kernels were replaced by rounding to the nearest experimental kernels, as in Eq. (10). After training, we tested the accuracy of the trained OCNN using the MNIST test dataset under various levels of simulated Gaussian noise on the activations. For each Gaussian noise level $\sigma$, we ran 7 noisy inference instances by randomly sampling $\varepsilon$ from $\mathcal{N}(0,\sigma^2)$ to obtain the mean and standard deviation of the accuracy.

### 3.3. Inference simulation of MCNN

Fig. 4 plots the inference accuracy at various noise levels, quantified by RMSE, for the MCNNs with our training method and that of BNN. The MCNN trained with our method can maintain the inference accuracy up to $\sigma$=0.5, where the accuracy is 75.0±3.2% for our method, and 47.3±3.1% for BNN training. These results show that adding random noise in the training has improved the performance in a mixed-signal scenario, an effect similar to the regularization on the neural network parameters [27]. Notice that here we trained the MCNN off-line by modeling the analog computation using random noise and deterministic error in the forward and backpropagation processes. An in-situ [14] forward pass through the physical MCNN setup can leverage the full potential of the speed and efficiency provided by the analog accelerator unit.

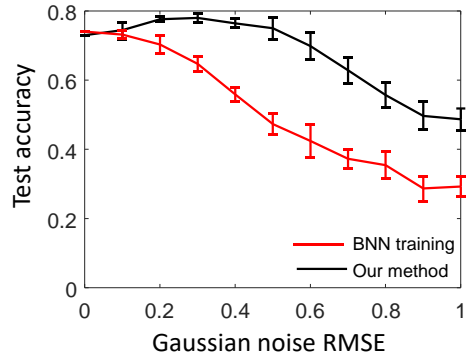

Figure 3. Classification accuracy as a function of noise RMSE added in the inference simulation.

Fig. 5 exemplifies the layer-by-layer activations of the two OCNNs in Fig. 4 at $\sigma$=0.5 for the input digit '5'. The probability of the input digit being classified as '5' is 99.2% and 83.1%, respectively, for the MCNN with our training method and that of BNN. Though the MCNN trained with BNN method still correctly classifies this digit, the probability is reduced to and confusions from digit '0', '3', and '8' can be observed from its output.

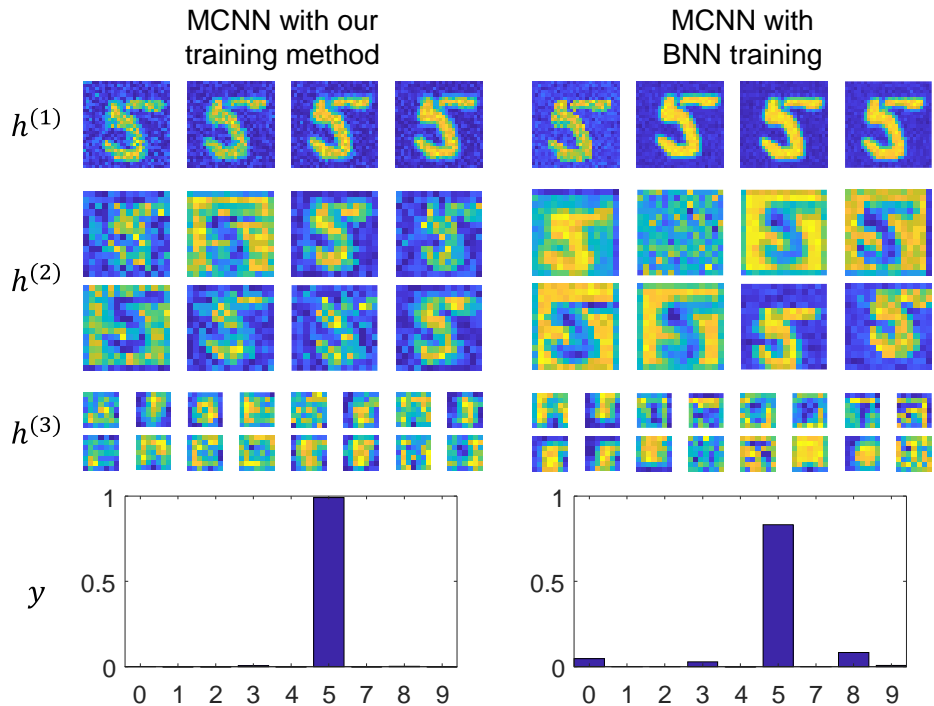

Figure 4. Layer-by-layer activations of the OCNNs trained with BNN and our method for the input digit 5.

Our treatment of the analog computation noise is similar to the stochastic quantization [10] to a lower precision level on a digital computer, indicating that mixed-signal neural network can operate at a reduced quantization level. Table 1 shows the inference accuracy of the two MCNNs on a simulated digital low-bit width system. We kept the same range of the activations, while stochastically quantizing them to 3, 2, and 1 bit(s), corresponding to 8, 4 and 2 quantization levels, respectively. For a binary input convolving with 3X3 binary kernels, the ideal activations range from 0 to 9. The fluctuation due to unbiased Gaussian noise with $\sigma$=0.5 can range from -1.0 to 1.0, considering the 95% confidence interval. The MCNN trained with our method maintains the accuracy at 2 bits (4 levels) or higher quantization levels, consistent with the amount of random noise it can tolerate.

Table 1: Inference accuracy of the OCNN with reduced activation quantization bit widths

| Quantization bit width | OCNN trained with our method | OCNN trained with standard backpropagation |
|---|---|---|
| 3 | 75.4±2.5% | 72.4±3.1% |
| 2 | 70.4±2.5% | 62.1±3.5% |
| 1 | 37.0±2.1% | 26.3±3.9% |

Quantization on digital, fixed-point neural networks is often performed stochastically to avoid introducing the quantization bias, which is undesirable in low-precision neural networks. Likewise, if there are some distortion uncorrected for in the training of MCNN, the residue deterministic error will introduce a bias, which is accumulative throughout all the layers. Fig. 6 plots the inference accuracy versus the RMSE of the residue error for the two MCNNs trained with our method and BNN method. The accuracy drop as the residue error increases is consistent with the results on digital platforms [8–10], indicating that our training is still sensitive to the bias from uncorrected errors.

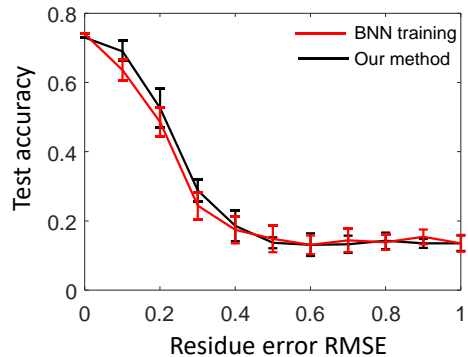

Figure 5. Accuracy vs. RMSE of residue deterministic errors added in the simulated inference process.

## 4. MCNN experiment

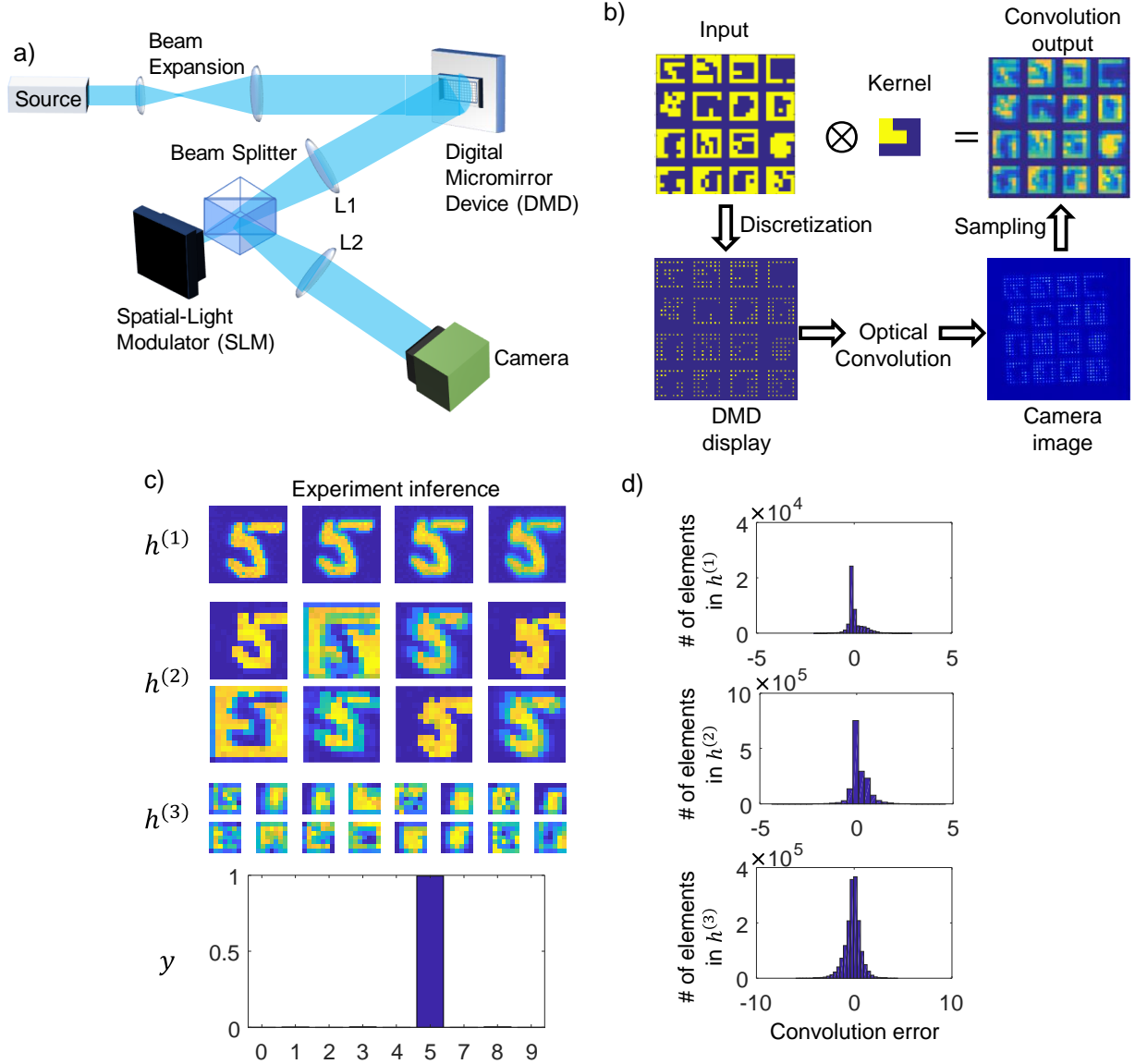

Figure 6. (a) Optical setup for implementing each MCNN layer. (b) Parallelized computation of the convolution between multiple input channels in layer 3 and the kernel. (c) Activations and output in each MCNN layer for the input digit '5'. (d) Histogram of the errors between ideal (convolution between DMD inputs and calibrated kernels) and experimental activations (down-sampled from raw camera images) in each layers.

Most of the existing diffractive optics-based neural networks employ non-erasable diffractive optical elements to represent a pre-trained set of weights [28,29], and hence cannot be re-programmed easily. Here, we constructed a fully programmable optical mixed-signal convolutional network layer based on a 4f system for the deployment of the trained MCNN model. The layer input is a digital mirror device (DMD, ViALUX, V4100 DLP7000, pixel size $13.7\ \mu m$). The analog convolution is performed by a phase-only spatial light modulator (SLM, Meadowlark Optics, P1920-400-800-HDMI, pixel size $9.2\ \mu m$) on the Fourier plane, as shown in Fig. 3(a). The DMD is illuminated by a collimated beam from a $12\ mW$ laser source (Coherent, OBIS LX $\lambda = 488nm$). Each element of each input channel, $\tilde{a}_{m,n}^{(k-1,q)}$, is represented by one DMD pixel, in either the on or the off state. The light field then passes through a $200mm$ tube lens (Thorlabs TTL-200), $L_1$, that creates a Fourier transform (FT) of the input onto the SLM. The FT of the kernel $\widetilde{W}^{(k,q,c)}$, approximated as phase only, is loaded onto the SLM. Upon reflection off the SLM, the FT of the input is multiplied by the FT of the kernel, thereby implementing the analog convolution. A beam splitter directs the reflected beam from the SLM through lens $L_2$ (identical to $L_1$), performing the inversion FT to yield the desired convolution between the input and the kernel, which is captured by a camera (JAI Ltd., GO-5000M-USB camera, 5.0 $\mu m$ pitch). To implement the CNN operation in Eq. (11), the kernels must be flipped horizontally and vertically before use.

The input patterns $\tilde{a}^{(k)}$ that can be displayed are strictly binary due to the use of the DMD as the input device, hence $\mathbf{A}' \in \{0,1\}$. The use of phases only masks [30] to approximate a complex Fourier filter gives rise to distortions in the kernels. For 3X3 binary kernels, there are a total of 511 non-trivial kernels. We pre-calculated the 511 phase masks needed to display all the non-trivial kernels. Because of the experimental artifacts and approximations, the actual 511 kernels, $\mathbf{W}'$, displayed in the experiment are not strictly binary. The distorted kernels are calibrated by imaging a single pixel displayed on the DMD through the optical system for each phase mask.

Due to aberrations and the limited numerical aperture of the 4f system, the full-width-at-half-maximum (FWHM) of its point-spread-function (PSF) is about 4 camera pixels. To mitigate crosstalk due to the PSF, we introduced a 3-pixel separation between adjacent samples of the input on the DMD and an 8-pixel separation for the kernels on the SLM, accounting for the differences between the pixel sizes of the DMD and the camera. To take advantage of the spatial bandwidth of the 4f system, we tiled multiple input channels in 2X2 and 4X4 formations for layer 2 and layer 3, respectively. Fig. 3(b) shows the tiled input on the DMD. After the raw image on the camera is acquired, we performed an 8X8 down sampling, and separate the tiled channels to recover $h$ in its native spatial resolution.

Fig. 3(c) shows the layer-by-layer activations of the MCNN model, trained with our method, in classifying the input digit '5'. With the calibrated kernel set $\mathbf{W}'$, ideal convolution between the DMD input and the actual kernels in each layer can be computed. We compared the ideal convolution results with the activations obtained from raw camera images. The distributions of errors in the activation tensor $h$ are plotted in Fig. 3(d), for all MCNN layers in the experiment. These distributions all resemble the Gaussian shape, with standard deviations $\sigma$=0.37, 0.38, and 0.58 for layer 1 through 3, respectively. The error in each layer is consistent with our choice of the random noise with $\sigma$=0.5 in the MCNN simulation. Despite the presence of this activation error, MCNN trained with our method achieved the correct inference.

## 5. Summary

We have demonstrated a training method to incorporate the analog computation error in neural network training for the deployment on mixed-signal computation platforms. Neural networks trained with our method is robust against a noise RMSE of 0.5 in the analog computing process, and thus can tolerate the reduced precision of the activations. Compared with a neural network trained using conventional backpropagation, our training method maintains the inference performance at approximately half of the precision levels determined by the data format and device specification. This allows us to deploy a trained convolutional neural network on a mixed-signal, diffractive optics-based convolution system that exhibits convolution error and kernel distortion.

# References

1. K. He, X. Zhang, S. Ren, and J. Sun, "Deep residual learning for image recognition," in *Proceedings of the IEEE Conference on Computer Vision and Pattern Recognition* (2016), pp. 770–778.

2. K. Simonyan and A. Zisserman, "Very deep convolutional networks for large-scale image recognition," arXiv Prepr. arXiv1409.1556 (2014).

3. C. Szegedy, W. Liu, Y. Jia, P. Sermanet, S. Reed, D. Anguelov, D. Erhan, V. Vanhoucke, and A. Rabinovich, "Going deeper with convolutions," in *Proceedings of the IEEE Conference on Computer Vision and Pattern Recognition* (2015), pp. 1–9.

4. A. Krizhevsky, I. Sutskever, and G. E. Hinton, "ImageNet Classification with Deep Convolutional Neural Networks," in *Advances in Neural Information Processing Systems 25*, F. Pereira, C. J. C. Burges, L. Bottou, and K. Q. Weinberger, eds. (Curran Associates, Inc., 2012), pp. 1097–1105.

5. R. Collobert and J. Weston, "A Unified Architecture for Natural Language Processing: Deep Neural Networks with Multitask Learning," in *Proceedings of the 25th International Conference on Machine Learning*, ICML '08 (Association for Computing Machinery, 2008), pp. 160–167.

6. D. Silver, T. Hubert, J. Schrittwieser, I. Antonoglou, T. Graepel, T. Lillicrap, K. Simonyan, D. Hassabis, A. Turing, and C. Shannon, "A general reinforcement learning algorithm that masters chess, shogi, and Go through self-play," Science (80-. ). **1144**, 1140–1144 (2018).

7. M. Rastegari, V. Ordonez, J. Redmon, and A. Farhadi, "XNOR-Net: ImageNet Classification Using Binary Convolutional Neural Networks BT - Computer Vision – ECCV 2016," in B. Leibe, J. Matas, N. Sebe, and M. Welling, eds. (Springer International Publishing, 2016), pp. 525–542.

8. I. Hubara, M. Courbariaux, D. Soudry, R. El-Yaniv, and Y. Bengio, "Quantized neural networks: Training neural networks with low precision weights and activations," J. Mach. Learn. Res. **18**, 1–30 (2018).

9. S. Zhou, Y. Wu, Z. Ni, X. Zhou, H. Wen, and Y. Zou, "DoReFa-Net: Training Low Bitwidth Convolutional Neural Networks with Low Bitwidth Gradients," arXiv Prepr. arXiv1606.06160 **1**, 1–13 (2016).

10. S. Gupta, A. Agrawal, K. Gopalakrishnan, and P. Narayanan, "Deep learning with limited numerical precision," 32nd Int. Conf. Mach. Learn. ICML 2015 **3**, 1737–1746 (2015).

11. A. Fan, P. Stock, B. Graham, E. Grave, R. Gribonval, H. Jegou, and A. Joulin, "Training with Quantization Noise for Extreme Model Compression," arXiv Prepr. arXiv2004.07320 1–18 (2020).

12. V. Sze, Y. Chen, T. Yang, and J. S. Emer, "Efficient Processing of Deep Neural Networks: A Tutorial and Survey," Proc. IEEE **105**, 2295–2329 (2017).

13. W. Haensch, T. Gokmen, and R. Puri, "The Next Generation of Deep Learning Hardware: Analog Computing," Proc. IEEE **107**, 108–122 (2019).

14. C. Li, D. Belkin, Y. Li, P. Yan, M. Hu, N. Ge, H. Jiang, E. Montgomery, P. Lin, Z. Wang, W. Song, J. P. Strachan, M. Barnell, Q. Wu, R. S. Williams, J. J. Yang, and Q. Xia, "Efficient and self-adaptive in-situ learning in multilayer memristor neural networks," Nat. Commun. **9**, 7–14 (2018).

15. P. Yao, H. Wu, B. Gao, J. Tang, Q. Zhang, W. Zhang, J. J. Yang, and H. Qian, "Fully hardware-implemented memristor convolutional neural network," Nature **577**, 641–646 (2020).

16. R. Hamerly, L. Bernstein, A. Sludds, M. Soljačić, and D. Englund, "Large-Scale Optical Neural Networks Based on Photoelectric Multiplication," Phys. Rev. X **9**, 1–12 (2019).

17. Y. Shen, N. C. Harris, S. Skirlo, M. Prabhu, T. Baehr-Jones, M. Hochberg, X. Sun, S. Zhao, H. Larochelle, D. Englund, and M. Soljačić, "Deep learning with coherent nanophotonic circuits," Nat. Photonics **11**, 441–446 (2017).

18. H. Tsai, S. Ambrogio, P. Narayanan, R. M. Shelby, and G. W. Burr, "Recent progress in analog memory-based accelerators for deep learning," J. Phys. D. Appl. Phys. **51**, (2018).

19. Y. Cai, T. Tang, L. Xia, M. Cheng, Z. Zhu, Y. Wang, and H. Yang, "Training low bitwidth convolutional neural network on RRAM," Proc. Asia South Pacific Des. Autom. Conf. ASP-DAC **2018**-**Janua**, 117–122 (2018).

20. S. Moon, K. Shin, and D. Jeon, "Enhancing Reliability of Analog Neural Network Processors," IEEE Trans. Very Large Scale Integr. Syst. **27**, 1455–1459 (2019).

21. M. Cheng, L. Xia, Z. Zhu, Y. Cai, Y. Xie, Y. Wang, and H. Yang, "TIME: A training-in-memory architecture for RRAM-based deep neural networks," IEEE Trans. Comput. Des. Integr. Circuits Syst. **38**, 834–847 (2019).

22. N. P. Jouppi, A. Borchers, R. Boyle, P. Cantin, C. Chao, C. Clark, J. Coriell, M. Daley, M. Dau, J. Dean, B. Gelb, C. Young, T. V. Ghaemmaghami, R. Gottipati, W. Gulland, R. Hagmann, C. R. Ho, D. Hogberg, J. Hu, R. Hundt, D. Hurt, J. Ibarz, N. Patil, A. Jaffey, A. Jaworski, A. Kaplan, H. Khaitan, D. Killebrew, A. Koch, N. Kumar, S. Lacy, J. Laudon, J. Law, D. Patterson, D. Le, C. Leary, Z. Liu, K. Lucke, A. Lundin, G. MacKean, A. Maggiore, M. Mahony, K. Miller, R. Nagarajan, G. Agrawal, R. Narayanaswami, R. Ni, K. Nix, T. Norrie, M. Omernick, N. Penukonda, A. Phelps, J. Ross, M. Ross, A. Salek, R. Bajwa, E. Samadiani, C. Severn, G. Sizikov, M. Snelham, J. Souter, D. Steinberg, A. Swing, M. Tan, G. Thorson, B. Tian, S. Bates, H. Toma, E. Tuttle, V. Vasudevan, R. Walter, W. Wang, E. Wilcox, D. H. Yoon, S. Bhatia, and N. Boden, "In-Datacenter Performance Analysis of a Tensor Processing Unit," in *Proceedings of the 44th Annual International Symposium on Computer Architecture - ISCA '17* (ACM Press, 2017), pp. 1–12.

23. J. Choi, Z. Wang, S. Venkataramani, P. I.-J. Chuang, V. Srinivasan, and K. Gopalakrishnan, "PACT: Parameterized Clipping Activation for Quantized Neural Networks," 1–15 (2018).

24. S. Jung, C. Son, S. Lee, J. Son, J. J. Han, Y. Kwak, S. J. Hwang, and C. Choi, "Learning to quantize deep networks by optimizing quantization intervals with task loss," Proc. IEEE Comput. Soc. Conf. Comput. Vis. Pattern Recognit. **2019**-**June**, 4345–4354 (2019).

25. B. E. A. Saleh and M. C. Teich, *Fundamentals of Photonics* (john Wiley & sons, 2019).

26. Y. Bengio, N. Léonard, and A. Courville, "Estimating or Propagating Gradients Through Stochastic Neurons for Conditional Computation," 1–12 (2013).

27. C. M. Bishop, "Training with Noise is Equivalent to Tikhonov Regularization," Neural Comput. **7**, 108–116 (1995).

28. J. Chang, V. Sitzmann, X. Dun, W. Heidrich, and G. Wetzstein, "Hybrid optical-electronic convolutional neural networks with optimized diffractive optics for image classification," Sci. Rep. **8**, 12324 (2018).

29. T. Yan, J. Wu, T. Zhou, H. Xie, F. Xu, J. Fan, L. Fang, X. Lin, and Q. Dai, "Fourier-space Diffractive

Deep Neural Network," Phys. Rev. Lett. **123**, 023901 (2019).

30. D. Mendlovic, G. Shabtay, U. Levi, Z. Zalevsky, and E. Marom, "Encoding technique for design of zero-order (on-axis) Fraunhofer computer-generated holograms," Appl. Opt. **36**, 8427 (1997).